\documentclass[aps,prc,twocolumn,superscriptaddress,groupedaddress]{revtex4}

\usepackage{dcolumn} 
\usepackage{hyperref}
\usepackage{color}
\usepackage{graphicx}

\usepackage{bm}        
\usepackage{amssymb}  

\begin{document}
\title{Onset of the ridge structure in AA, pA and pp collisions}
\author{C. Andr\'es}
\author{A. Moscoso}
\author{C. Pajares}
\affiliation{Departamento de F\'isica de Part\'iculas and IGFAE, Universidade de Santiago de Compostela, 15782, Santiago de Compostela, Spain}

\begin{abstract}
It is shown that the anomalous sharp increasing of the strength of the near-side ridge structures observed in Au-Au collisions at $\sqrt{s}=$ 62 GeV and $\sqrt{s}=$ 200 GeV and the onset of the ridge structure in pPb and in pp collisions can be naturally explained in the framework of string percolation. In all the cases the near-side strength reflects the collision area covered by the strings stretched between the colliding objects and therefore it is related to the shape of their profile functions. The dependence of the pseudorapidty and azimuthal widths on multiplicty and energy is qualitatively explained.
\end{abstract}
\maketitle
\section{Introduction}

Correlations between pairs of hadrons that are collimated in their relative azimuthal angle and are long range in relative rapidity were observed in heavy ion collisions at energies available at the BNL Relativistic Heavy Ion Collider (RHIC) \cite{ref1,ref1b,ref2,ref3,ref4,ref4b} and later at  the CERN Large Hadron Collider (LHC) energies \cite{ref31}. These ridge-like correlations have also been seen in proton-proton collisions at $\sqrt{s} =$ 7 TeV for high multiplicity events \cite{ref5}. More recently, a sizable ridge has been seen in p-Pb collisions at $\sqrt{s} =$ 5.02 TeV \cite{ref6,ref7,ref8,ref32,ref33}. Much attention has been paid to understand whether these structures are due to initial state or to final state effects that are amenable to a hydrodynamic description \cite{ref9,ref10,ref11,ref12,ref13,ref14,ref15,ref16,ref17,ref17b,ref18,ref19,ref20,ref21}. 

The origin of long range rapidity correlations is similar in heavy ion and proton-proton collisions and it is explained, in the glasma picture of the Color Glass Condensate (CGC), due to the saturation of color flux tubes correlated in the transverse space with a length 1/$Q_{s}$ determined by the saturation momentum $Q_{s}$ \cite{ref13,ref14,ref15,ref16,ref17,ref17b,ref18}. On the contrary, not much attention has been paid to the onset of the ridge structure. In pp collisions at $\sqrt{s} =$ 7 TeV the structure is only observed for N$_{ch} >$ 110 and in pPb collisions at $\sqrt{s} =$ 5.02 TeV for N$_{ch} \gtrsim$ 50 ($N_{ch}$ is the number of charged particles created in a particular collision). Moreover, in Au-Au collisions at $\sqrt{s} =$ 200 GeV and $\sqrt{s} =$ 62 GeV an anomalous centrality dependence of the correlation is observed: the strength of the near-side ridge as a function of multiplicity presents a change on the behavior of the slope for $N_{ch} =$ 120 and $N_{ch} =$ 130 respectively \cite{ref22}.

In this paper, we show that all these features can be understood in the framework of percolation of strings. As the number of strings formed in a collision reaches an universal critical density, a macroscopic cluster of strings appears covering around $2/3$ of the total collision area \cite{ref23}. As the density approaches the critical value the ridge structure begins to unfold. The dependence of the strength of the near-side ridge on the multiplicity in a given collision reflects the fraction of the collision area covered by strings which is related to the profile function of the colliding objects.

Although the goal of this paper is not to give a detailed description of the azimuthal dependence of the ridge structure, we include a brief discussion on how the collimated $\phi$-distribution of the near-side ridge arises in our approach.

\section{The near-side ridge in the string percolation model}

The strings are the basic ingredients of the model and it is necessary to know their number, rapidity extension, fragmentation and probability distribution, which depend on the chosen model. This model dependence is not strong because in most of the color exchange string models, as dual parton model (DPM) \cite{ref34,ref35}, quark-gluon string model \cite{ref36}, VENUS \cite{ref37} or EPOS \cite{ref38}, the results for the mentioned observables are very similar. In the present study we use the DPM. In this model, 2$k$ strings are produced in pp collisions, two of them stretched between a valence diquark of the projectile (target) and a valence quark of the target (projectile) and 2$k-$2 strings stretched between sea quarks and antiquarks. Due to the momentum distribution functions of the valence diquarks, valence quarks and sea quarks or antiquarks, $x^{3/2}$, $x^{-1/2}$ and $x^{-1}$ respectively, the rapidity extension of the $(qq)_v-q_v$ strings is in between one edge of the rapidity range and the central rapidity region. On the contrary the $q_s-\bar{q}_s$ strings are in the central rapidity region which approximately grows in rapidity extension proportional to the longitudinal phase space, $\ln s$. The mean number of strings at a given energy is determined by the cross section of producing 2$k$ strings, $\sigma_{2k}$, corresponding to cut $k$ Pomerons. At not very high energy, there are only contributions from the two $(qq)_v-q_v$ strings. As the energy increases, there are more and more contributions of the $q_s-\bar{q}_s$ type. In the case of AA collisions, the rapidity distribution is approximately given by \cite{ref35}
\begin{eqnarray}
\label{p2}
\frac{dN}{dy}&&\approx \left<N_A\right>\left(2N^{(qq)_v-q_v}(y)+2(<k>-2)N^{q_s-\bar{q}_s}(y)\right)  \nonumber \\ &&+\left(\left<N_C\right>-\left<N_A\right>\right)2<k>N^{q_s-\bar{q}_s}(y)\,,
\end{eqnarray}
where $\left<N_A\right>$ and $\left<N_C\right>$ are the mean number of participants and collisions, obtained in the Glauber-Gribov model \cite{refp1}. $N^{(qq)_v-q_v}(y)$ and $N^{q_s-\bar{q}_s}(y)$ are the rapidity distributions of the corresponding $(qq)_v-q_v$ and $q_s-\bar{q}_s$ strings. 

The strings decay into new ones by $q-\bar{q}$ and $qq-\bar{q}\bar{q}$ pair production and subsequently hadronize producing the observed particles. Due to confinement the color of these strings is confined to a small area in the transverse plane, $S_1=\pi r_0^2$. According to lattice and field correlator studies \cite{refp2,refp3} $r_0\approx$ 0.2 fm. 

With increasing energy and/or atomic number of the colliding particles, the number of strings grows and they start to overlap forming clusters, very similar to discs in two-dimensional percolation theory. Defining the density of strings as $\rho = N_s\frac{S_1}{S_A}$, where $S_A$ is the collision area and $N_s$ the number of strings, at a critical density $\rho_c =$ 1.2-1.5 a macroscopical cluster appears, which marks the percolation phase transition. The value of $\rho_c$ varies between 1.2 and 1.5 depending on the profile used (1.2 in the homogeneous case and 1.5 for the Gaussian or Wood-Saxon type) \cite{ref27}.

A cluster of $n$ strings behaves as a single string with energy-momentum corresponding to the sum of the individual ones and with a higher color field corresponding to the vectorial sum in color space of the color fields of the individual strings. Due to the randomness of the color field, the strength of the resulting color field is not $n$ times the strength of the individual color field but $\sqrt{n}$. Due to this fact, the multiplicity and transverse momentum of the clusters are given by \cite{ref24,ref25}:
\begin{equation}
\mu_n=\sqrt{n\frac{S_n}{S_1}}\mu_1, \quad \langle p^2_T\rangle_n=\sqrt{n\frac{S_1}{S_n}}\langle p^2_T\rangle_1,
\label{eq1}
\end{equation}
where $\mu_1$ and $\langle p^2_T\rangle_1$ are, respectively, the multiplicity and the $p^2_T$ of the particles produced by a single string and $S_n$ is the area covered by $n$ strings. In the high density limit the formula (\ref{eq1}) can be written as \cite{ref24}:
\begin{equation}
\label{eq2}
\mu_n=N_sF(\rho)\mu_1, \quad \langle p^2_T\rangle_n= \langle p^2_T\rangle_1/F(\rho),
\end{equation}
where
\begin{equation}
F(\rho)=\sqrt{\frac{1-e^{-\rho}}{\rho}}.
\label{eq3}
\end{equation}

Assuming a homogeneus profile for the collision area, the distribution of the overlapping of $n$ strings is Poissonian with a mean value $\rho$, $P_n = \frac{\rho^n}{n!}e^{-\rho}$. Hence, in (\ref{eq3}), $1-e^{-\rho}$ is the fraction of the area covered by strings. A more realistic profile implies a modification of the area covered by strings in formula (\ref{eq3}).

The area covered by clusters divided by the area of an effective cluster gives the effective average number $\langle N\rangle$ of clusters
\begin{equation}
\langle N\rangle=\frac{(1-e^{-\rho})R^2}{r^2_0F(\rho)}=\sqrt{1-e^{-\rho}}\sqrt{\rho}\left(\frac{R}{r_0}\right)^2\,,
\label{eq4}
\end{equation}
where R is the radius of the collision area (for non central collisions it should be used the corresponding area $S$ of the almond shape. In this case, instead of $\left(R/r_0\right)^2$ it should be $S/S_1$). 

The energy-momentum of the cluster of strings is the sum of the energy-momentum of the individual strings. As in the central rapidity region the main contribution, according to the Eq. (\ref{p2}), comes from the $q_s-\bar{q}_s$ strings, whose number is proportional to $(N_C-N_A)$, the mean rapidity extension of one string should be
\begin{equation}
\label{eq6}
\Delta y_1=c_1\left(1-\frac{N_A}{N_C}\right)\ln\left(\frac{s}{s_0}\right)\,,
\end{equation}
where $s_0$ is the minimum energy required for the creation of a single string able to decay in two particles. We take $s_0=$1 $\mathrm{GeV}^2$.

On the other hand, as each cluster contains on average $\frac{N_s}{N}=\frac{1}{F(\rho)}$ strings, the rapidity length of these effective clusters is
\begin{eqnarray}
\label{eq5}
\sigma_{\Delta y}=\Delta y_1-\ln F(\rho)&&=c_1\left(1-\frac{N_A}{N_C}\right)\ln\left(\frac{s}{s_0}\right)\nonumber\\&&+\ln\sqrt{\frac{\rho}{1-e^{-\rho}}}\,.
\end{eqnarray}
This is the rapidity width of the near-side ridge. The constant $c_1$ is independent of $N_A$, $N_C$ and energy. 

The normalized two-particle correlation function can be written in the two step scenario \cite{ref24,ref25} as the normalized fluctuation in the number of effective strings

\begin{equation}
\label{eq7}
\mathcal{R}\equiv \frac{\langle n^2\rangle-\langle n\rangle^2-\langle n\rangle}{\langle n\rangle^2}=\frac{\langle N^2\rangle-\langle N\rangle^2}{\langle N^2\rangle}=\frac{1}{k},
\end{equation}
where $n$ is the multiplicity of the produced particles. If the particle distribution is a negative binomial, as in the case of string percolation, with parameter $K_{NB}$ then $k = K_{NB}$. In the low density regime, $\rho$ small, the multiplicity distribution is essentially Poisson-like and therefore $k\rightarrow\infty$. In the large density limit, assuming that the $N$ effective clusters behave like a single one, $\langle N^2\rangle-\langle N\rangle^2 \approx \langle N\rangle$, then $k\rightarrow\langle N\rangle\rightarrow\infty$ \cite{ref25,ref26}. At intermediate string density, $k$ must have a minimum close to the critical string density. A parametrization of $k$ satisfying the above requirements, having a minimun close to the critical density, $\rho_c=1.2$, is \cite{ref20}

\begin{equation}
\label{eq8}
k=\frac{\sqrt{\rho}\left(R/r_0\right)^2}{1-e^{-\rho}}=\frac{\langle N\rangle}{\left(1-e^{-\rho}\right)^{3/2}}.
\end{equation}
Any other possible parametrization with the above requirements can not be very different from Eq. (\ref{eq8}) in the $\rho$ range around the critical value. 

The experimental data on Au-Au at 62 GeV and 200 GeV measures $\frac{\Delta\rho}{\sqrt{\rho_{ref}}}$, where $\Delta\rho$ is the difference between the pair distribution of the same collision and the uncorrelated pair distribution $\rho_{ref}$ obtained from mixed events. It is shown (appendix C of reference \cite{ref22}) that this quantity is proportional to $\mathcal{R}\frac{dn}{dy}$. The STAR collaboration fit the data on the near side ridge structure, obtaining the value $A_1$ of its strength for different centralities at the two different considered energies of the experiment. Hence, we write
\begin{equation}
\label{eq9}
\frac{\Delta \rho}{\sqrt{\rho_{ref}}}=\mathcal{R}\frac{dn}{dy}G(\phi)\,,
\end{equation}
where $G(\phi)$ encodes all the azimuthal dependence. This factorized form was used before in the framework of the Color Glass Condensate \cite{ref7,ref13b}.

From equations (\ref{eq4}), (\ref{eq5}), (\ref{eq7}) and (\ref{eq8}) we deduce
\begin{equation}
\label{eq10}
\mathcal{R}\frac{dn}{dy}=\frac{\langle N\rangle}{k}=\left(1-e^{-\rho}\right)^{3/2}\,.
\end{equation}
As previously mentioned, the factor $1-e^{-\rho}$ is the fraction of the collision surface covered by strings, assuming a homogeneus profile \cite{ref24}. For more realistic profiles this shape is modified. For Gaussian and Wood-Saxon profiles, the critical density is not $\rho_c\approx$1.2 but $\rho_c\approx$1.5 and the fraction of the collision area covered by strings is more similar to \cite{ref27}
\begin{equation}
\label{eq13}
A(\rho)=\frac{1}{1+ae^{-(\rho-\rho_c)/b}}\,,
\end{equation}
where $\rho_c=$1.5. The parameter $b$ controls the ratio between the edge $(2\pi R)$ and the total surface $(\pi R^2)$ and therefore is proportional to the inverse of the radius. Therefore, instead of using $1-e^{-\rho}$ in Eq. (\ref{eq10}) we use $A(\rho)$ and from Eq. (\ref{eq9}) we have
\begin{equation}
A_1=cA^{3/2}(\rho)\,.
\end{equation}

The main aim of this paper is not a detailed study of the azimuthal dependence of the near-side ridge. However, a brief discussion on how a collimated $\phi$-distribution arises in the string percolation framework is in order. The conventional understanding of the ridge is simply related to flow harmonics in a fluid dynamic scenario, where the inclusion of the pp and pA ridges is a challenge. In string percolation, the fragmentation of a single string gives a rather flat $\phi$-distribution. In the $3$-dimensional space the string becomes non orthogonal to the transverse plane and one may expect an anistropic emission of particles in this plane but this anisotropy is very small \cite{ref42}. However, a cluster of overlapping strings has an asymmetric form in the transverse plane and the partons emitted at the same point inside the cluster have to pass through a certain length before appearing outside and being observed. Along this path, they interact with the strong color field inside the cluster and their energy decreases. As a result, an observed particle with transverse momentum $p_T$ was born inside the cluster with a higher momentum whose value depends on the path legth travelled inside the cluster. Consequently, this moment is different for the different directions of emission. As the trasnverse distribution of the strings depends on the form of the nuclear ovelapping, so do the azimuthal distribution of strings and of the cluster of strings. These two anisotropies lead to the anisotropy of the spectrum of emitted particles. In this way, a reasonable agreement is obtained with the experimental data on $v_2$ \cite{ref42} and higher harmonics \cite{ref43}.

The narrow structure in the azimuthal dependence of the near-side ridge is determined by the transverse correlation length which, according to equation (\ref{eq2}), is $r_0F(\rho)^{1/2}$. Hence, the width of the azimuthal angle is
\begin{equation}
\label{eq17}
\sigma_{\Delta \phi}=c_2\left(\frac{1-e^{-\rho}}{\rho}\right)^{1/4}.
\end{equation}

Eq. (\ref{eq17}) is only a rough evaluation of the width of the ridge structure and therefore the comparison with data should be considered carefully. In order to do a more detailed comparison a Monte-Carlo simulation is necessary. The first results of this simulation agree with the experimental data on the near side width for pp, pPb and Au-Au collisions \cite{ref44}.

\section{Comparison with experimental data (RHIC, LHC)}

\begin{figure}
\includegraphics[scale=0.45]{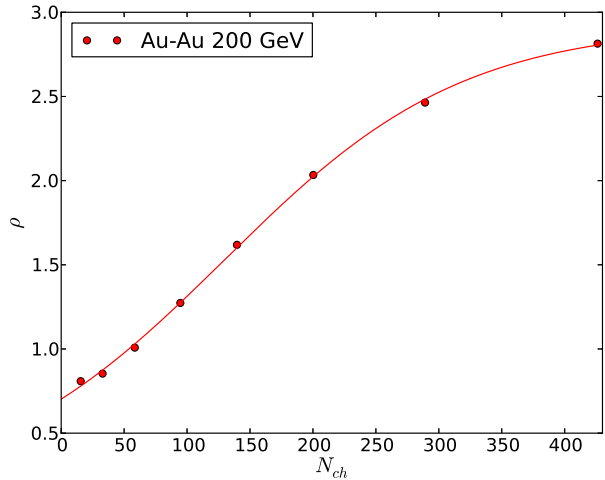}
\caption{(Color online.) String density $\rho$ plotted as a function of $N_{ch}$ for Au-Au collisions at $\sqrt{s}=$ 200 GeV.}
\label{fig2}
\end{figure}

In order to compare with the experimental data on the strength of the near-side ridge we use equation (\ref{eq13}), fitting the expression $A_{1} = cA(\rho)^{3/2}$ to the data on Au-Au collisions at $\sqrt{s}=$ 200 GeV and $\sqrt{s}=$ 62 GeV and pPb and pp collisions at $\sqrt{s}=$ 5.02 TeV and $\sqrt{s}=$ 7 TeV respectively. To do this we need to know the values of $\rho$. They are taken from a previous study in the framework of string percolation of $dN/dy$ at SPS, RHIC and LHC energies for pp and AA at different centralities \cite{ref28,ref29}. In the case of pp and pPb to compute $\rho$ we use the collision area of references \cite{ref18} and \cite{ref19}. The values of $\rho$ for Au-Au collisions at $\sqrt{s}=$ 200 GeV for the different centralities are in the range 0.6-3.0. For pPb at $\sqrt{s}=$ 5.02 TeV the values of $\rho$ are between 0.8-2.3 corresponding to $N_{ch} =$ 50 and to $N_{ch}=$ 330, respectively, and for pp at $\sqrt{s}=$ 7 TeV the values are in the range 0.25-0.7, corresponding to N$_{ch} =$ 18 (minumum bias) and N$_{ch} =$ 110, respectively. In high multiplicity pp collisions at LHC the values of $\rho$ are close to the values obtained in peripheral Cu-Cu collisions at RHIC energies, where a ridge structure was observed. This fact was the main reason to predict \cite{ref11} the near-side ridge later observed. An example of the dependence of $\rho$ on centrality is shown in Fig.~\ref{fig2} for Au-Au collisions at $\sqrt{s}=$ 200 GeV.

\begin{figure}
\includegraphics[scale=0.45]{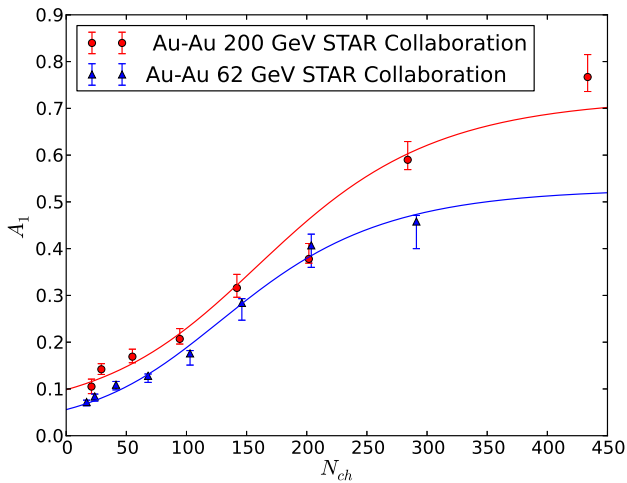}
\caption{(Color online.) Comparison between our results on the strength of the near-side ridge for Au-Au collisions at two RHIC energies $\sqrt{s}=$ 200 GeV (red line) and $\sqrt{s}=$ 62 GeV (blue line) with experimental data \cite{ref22} in terms of $N_{ch}$.}
\label{fig3}
\end{figure}

\begin{figure}
\includegraphics[scale=0.45]{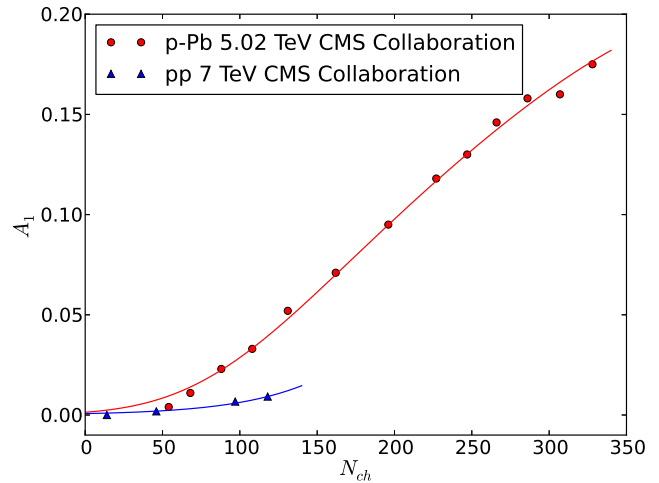}
\caption{(Color online.) Comparison between our results on the strength of the near-side ridge for pp collisions at $\sqrt{s}=$ 7 TeV (blue line) and pPb collisions $\sqrt{s}=$ 5.02 TeV (red line) with experimental data \cite{ref6} versus $N_{ch}$.}
\label{fig4}
\end{figure}

In Fig.~\ref{fig3} we show our results on the strength, $A_1$, of the near-side ridge as a function of the charged particles for Au-Au collisions at $\sqrt{s}=$ 200 GeV and $\sqrt{s}=$ 62 GeV. The values of the parameters obtained are $a =$ 1.5 and $b=$ 0.75. In Fig.~\ref{fig4} we show our results for pPb and pp. In this case, the parameters are $a =$ 1.5 and $b =$ 0.35. The values of $a$, $b$ and $c$ for the different collisions are summarized in Table \ref{tab1}.
\begin{table}
\begin{tabular} { c | c | c | c | c }
 & Au-Au 200 GeV & Au-Au 62 GeV & pPb & pp \\ \hline
 a & 1.5 &  1.5 &  1.5 &  1.5   \\ \hline
 b & 0.75 & 0.75 & 0.35 & 0.35 \\ \hline
 c & 0.93 & 0.80 & 0.21 & 0.57 \\
\end{tabular}
\caption{Values of parameters $a$, $b$ and $c$ for the different collisions.\label{tab1}}
\end{table}
We obseve that the value of $b$ in the pPb and pp case is much smaller than in the case of Au-Au collisions as it was expected. Notice that the string density in pPb at $N_{ch} =$ 50, where the near-side ridge structure emerges, is $\rho =$ 0.8. In the pp case the near-side ridge structure unfolds at $N_{ch} =$ 100, corresponding to $\rho =$ 0.7, very close to the obtained one for pPb collisions. Notice that, apart from the normalization constant $c$, different for the three type of collisions, we fix the parameters $\rho_c =$ 1.5 and $a =$ 1.5 for all the considered collisions, keeping $b$ as the only fitting parameter. Even this parameter is not fully free because its dependence on the nucleus radius should be similar to $1/R$, as it is obtained.

\begin{figure}
\includegraphics[scale=0.45]{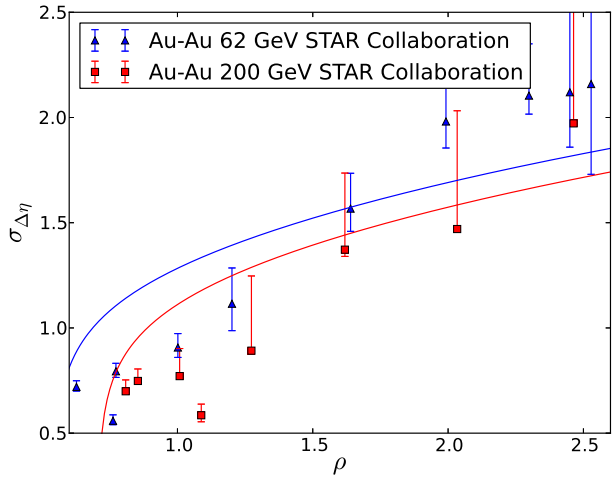}
\caption{(Color online.) Pseudorapidity width of the near-side ridge for Au-Au collisions at two RHIC energies $\sqrt{s}=$ 200 GeV and $\sqrt{s}=$ 62 GeV \cite{ref22} plotted as a function of $\rho$. Curves obtained from equation (\ref{eq5}) for Au-Au collisions at $\sqrt{s}=$ 200 GeV (red) and $\sqrt{s}=$ 62 GeV (blue).}
\label{fig5}
\end{figure}

In Fig.~\ref{fig5} we compare our results on the pseudorapidity width, obtained from equation (\ref{eq5}) with the experimental results on Au-Au at $\sqrt{s}=$ 200 GeV and $\sqrt{s}=$ 62 GeV \cite{ref22}. The value of $c_1$ is 0.23. $N_A$ and $N_C$ are taken from the quoted experimental analysis \cite{ref30}. It is observed that our result for $\sqrt{s}=$ 200 GeV is slightly larger than the corresponding one at $\sqrt{s}=$ 62 GeV. Experimental data are very close at both energies.

\begin{figure}
\includegraphics[scale=0.45]{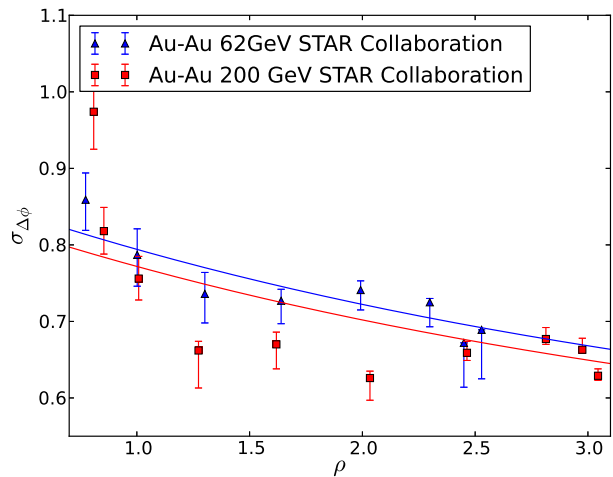}
\caption{(Color online.) Azimuthal width of the near-side ridge for Au-Au collisions at $\sqrt{s}=$ 200 GeV and $\sqrt{s}=$ 62 GeV \cite{ref22} versus $\rho$. Curves obtained from equation (\ref{eq17}) for Au-Au collisions at $\sqrt{s}=$ 200 GeV (red) and $\sqrt{s}=$ 62 GeV (blue).}
\label{fig6}
\end{figure}

In Fig.~\ref{fig6} experimental data on the azimuthal width for Au-Au collisions at $\sqrt{s}=$ 200 GeV and $\sqrt{s}=$ 62 GeV \cite{ref22} are compared to our model. The values of $c_2$ are $c_2 =$  0.866 for Au-Au at $\sqrt{s}=$ 200 GeV and $c_2 = 0.890$ for Au-Au at $\sqrt{s}=$ 62 GeV. The azimuthal width decreases with increasing energy and centrality in agreement with the trend of the experimental data. For both widths a qualitative agreement is obtained.

The dependence of A$_1$, $\sigma_{\Delta \eta}$ and $\sigma_{\Delta \phi}$ on energy and centrality, resulting from equations (\ref{eq10}) and (\ref{eq13}), is very similar to the obtained one in the glasma picture. In this approach, $\mathcal{R}dN/dy$ and $\sigma_{\Delta \eta}$ are proportional to 1/$\alpha_s(Q_s$) and $\sigma_{\Delta \phi}$ is proportional to $1/Q_s$. Hence, both $\mathcal{R}dN/dy$ and $\sigma_{\Delta \eta}$ grow with $\ln s$ and $\ln N_A$. In the high density limit, we obtain in percolation the same dependence on $s$ and on N$_A$ for both observables. In the case of $\sigma_{\Delta \phi}$, as 1/$Q_s$$\sim N_A^{1/6}s^{\Delta/2}$ and $r_0F(\rho)^{1/2}\sim r_0\rho^{1/4} \sim r_0N_A^{1/6}s^{\Delta/2}$.

\section{Conclusions}

In conclusion, we have shown that percolation of strings naturally explains the anomalous dependence of the near-side ridge structure correlation on the multiplicity observed in Au-Au collisions at $\sqrt{s}=$ 200 GeV and $\sqrt{s}=$ 62 GeV. The onset of the ridge structure in high multiplicity pp events at $\sqrt{s}=$ 7 TeV and in high multiplicity pPb at $\sqrt{s}=$ 5.02 TeV are also explained. Furthermore, our model qualitatively describes the dependence of the azimuthal and pseudorapidity widths on multiplicity. Our framework can be regarded as a complementary picture to the glasma in the description of the initial state, able to explore the transition from low to high density. Most of the ingredients used can be considered initial state effects, although the quenching of the partons produced in the cluster of strings is a final state effect.  

\section*{Acknowledgements}

We thank G. Feofilov, M. A. Braun, N. Armesto and C. Salgado for very useful discussions. This work has been done under the project FPA2011-22776 of MINECO (Spain), the Spanish Consolider CPAN project, FEDER funds and Xunta de Galicia (GRC 2013-024).

\bibliographystyle{apsrev}
\bibliography{manuscript}
\end{document}